\documentclass{article}
\usepackage{hiph-preprint}
\usepackage{epsfig}
\volnumber{22} \issuenumber{1} \edyear{2005}                             
\frompage{000} \topage{000}                                              
\recrevdate{1 January 2005}                                              
\def\rightmark{Charge Correlation in H-H Jets}
\def\leftmark{A. Kravitz for the PHENIX Collaboration}

\newcommand{\Au} {\mbox{$Au-Au$}}

\newcommand{\snn} {\mbox{$\sqrt{s_{NN}}$}}

\newcommand{\pt} {\mbox{$p_T$}}
\newcommand{\vtwo} {\mbox{$v_2$}}
\newcommand{\dphi} {\mbox{$\Delta\phi$}}

\title{Charge Correlation in Near Side Hadron-Hadron Jets at \snn\ = 200 GeV in PHENIX}
\authors{
{Alan Kravitz$^1$ (for the PHENIX Collaboration) %
\index{Kravitz, A.} 
}\\[2.812mm]
{\normalsize
\hspace*{-8pt}$^1$ Columbia University,\\
New York, USA 10027 }}

\abstract{We analyze the relative azimuthal (\dphi) distribution of
same-charge and opposite-charge particle pairs. We then remove
elliptic flow background using the ZYAM method. Comparisons between
near angle \dphi\ peak widths are presented for various
centralities.}



\makeindex
\begin{document}

\maketitle

\section{Introduction}\label{intro}
In high energy heavy ion collisions at RHIC, the dihadron
correlation function is known to be strongly modified by the medium.
These modifications are reflected in changes in the \pt\ spectrum,
flavor composition, shape and yield of the jet. It has been
shown\cite{chrgassym} that there is a charge-asymmetry within jets
that favors oppositely charged particle pairs. By splitting the
correlation function by charge-pair, we can better study both the
jet signal and detector effects.

\section{Two Particle Azimuthal Correlation}\label{techno}

\subsection{The Correlation Function}\label{ss_corrfunc}
\begin{figure*}[htb]
\begin{center}
\epsfig{file=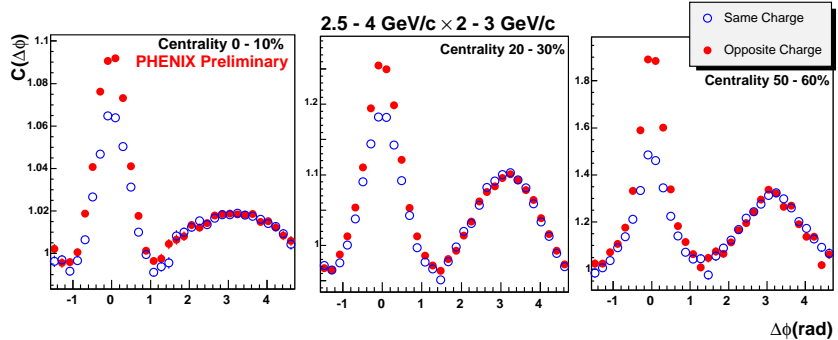,width=0.9\linewidth}
\vspace{-.5cm}
 \caption{\label{corrfunc} The \dphi\ correlation
functions for various centralities.}
\end{center}
\end{figure*}

This analysis is based on 1 billion \Au\ events from the PHENIX run
4 dataset at \snn\ = 200 GeV. We construct the correlation function
($C(\Delta\phi)$) using trigger particles between 2.5-4 GeV/c,
correlated with associated particles with momentum between 2-3
GeV/c. The correlation function is represented by the following
formula:
\begin{eqnarray}
C(\Delta\phi)=J(\Delta\phi) + \xi(1+2v_{2, trig}v_{2,
assoc}cos(2\Delta\phi))
\label{corr_eq}
\end{eqnarray}
where $J(\Delta\phi)$ is the jet signal and  $v_{2, trig}$ and
$v_{2, assoc}$ are the eliptic flow for the trigger and associated
particles, respectively. $\xi$ is the normalization factor. Fig
\ref{corrfunc} shows the correlation function of same-sign and
opposite-sign pairs for three centrality bins.

On the near side, the magnitude of the same-charge pair distribution
is consistently smaller than the opposite-pair distribution. The two
distributions show differences that persist until around $\pm$1.2
rad. The away-side distributions agree with each other because the
trigger particles and the associated particles come from different
jets, and have no intrinsic charge correlation.

\subsection{The Charge-Pair Dependence of the Jet Shape}\label{ss_jet}

\begin{figure*}[htb]
\begin{center}
\epsfig{file=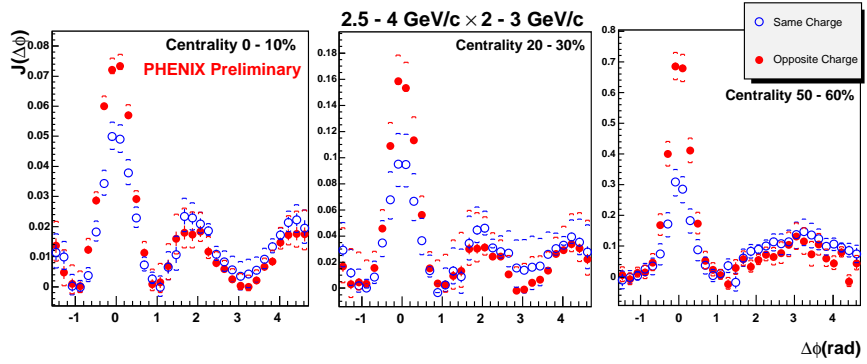,width=0.9\linewidth}
\vspace{-0.5cm}
 \caption{\label{corrfunc_v2corr} The \vtwo\ corrected (using ZYAM) signal for various centralities.
}
\end{center}
\end{figure*}

We then subtract the eliptic flow background from the correlation
function using the ``Zero Yield at Minimu'' (ZYAM) method\cite{ZYAM}
($J(\Delta\phi_{zyam})=0$). Figure \ref{corrfunc_v2corr} shows the
results. The systematic errors shown are dominated by uncertainties
in \vtwo, which are correlated between same-sign and opposite-sign
pairs. Note that the points on the away-side are no longer directly
on top of each other. This is because even at the minimum, the jet
still produces some yield. This overestimation of the elliptic flow
background results in more over-subtraction in the opposite-sign
pairs (where the jet signal is greater) than in the same-sign pairs.

\begin{figure*}[htb]
\begin{center}
\epsfig{file=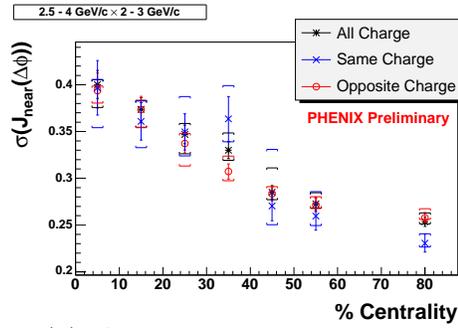,width=0.5\linewidth}
\vspace{-0.5cm}
 \caption{\label{v2corr_sigma} The jet width ($\sigma$)
 of the opposite charge, same charge correlation function for various centralities.}
\end{center}

\end{figure*}

The rest of this discussion focuses on the near-side jet shape.
In figure \ref{v2corr_sigma} we present the near side jet width for the
same-charge, opposite-charge and min bias pairs, obtained by fitting
a gaussian function over $\pm$ 1.2 radians. The stability of the fit
is evaluated by varying the fit range from $\pm$ 0.7 radians to
$\pm$ 1.5 radians and by fitting the entire distribution to a triple
gaussian formula (the two additional gaussians are used to describe
the split in the away-side distribution\cite{volcano}). The
systematic error shown is the maximum/minimum width obtained. The
width of the same-charge, opposite-charge, and min-bias pairs agree
with each other within errors. This demonstrates that the errors
inherent in the ZYAM assumption (that $J(\phi_{zyam})=0$) do not
significantly change the shape of the near-side jet.


There are a number of possible causes for the broadening of the jet
in most central collisions. It could be that the jet is broadened as
it traverses the medium. The more medium to traverse, the greater
the broadening. It is also possible that this broadening is related
to the decrease of per-trigger baryon yields vs. meson yields in
central collisions\cite{barmes}. We know that at higher centrality,
more baryons are produced, so this could just be an effect of the
jet composition and not of the medium.

We can subtract the two distributions in Fig. \ref{corrfunc} without
changing the width of the near side distribution. The \vtwo\ term
and the away-side jet are not dependent on charge sign, so they will
be subtracted out. The only part that will remain is the near-side
jet difference.

\begin{eqnarray}
\Delta J(\Delta\phi) = C_{opp}(\Delta\phi)-C_{same}(\Delta\phi) =
J_{opp}(\Delta\phi) - J_{same}(\Delta\phi)
\end{eqnarray}

\begin{figure*}[htb]
\begin{center}
\epsfig{file=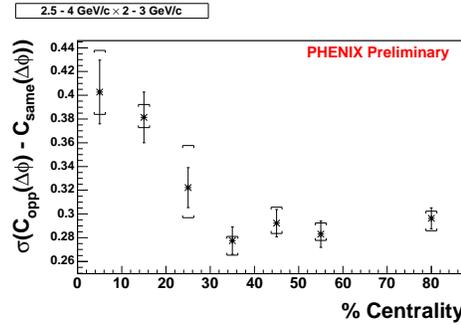,width=0.5\linewidth}
\vspace{-0.5cm}
 \caption{\label{corrfunc_diff_sigma} The jet width ($\sigma$) of
 $\Delta J$ for various centralities.}
\end{center}
\end{figure*}

This jet difference is fitted to a single gaussian centered at zero
over a range of $\pm$ 1.5 radians. Systematic errors on the width
of the near side jet are generated by varying the fit range between
$\pm$ 1 and $\pm$ 3 radians. The resulting jet width for different
centrality bins is presented in Fig \ref{corrfunc_diff_sigma}. The
distribution once again shows a decrease in the jet width in more
central collisions, consistent with Fig \ref{v2corr_sigma}.

\section{Summary}\label{concl}
Analysis of the same-charge and opposite-charge two particle
azimuthal correlations are very interesting because both \vtwo\ and
the away-side jet are both charge independent. The opposite-charge
correlation function shows a larger jet signal than the same-charge
correlation function, relative to the eliptic flow background. We
show that the jet width broadens in the most central collisions for
both the opposite-charge and same-charge pair distributions.

\vfill\eject
\end{document}